\documentclass[nouppercase]{cmbec}

\usepackage{url}
\usepackage[colorlinks=true, allcolors=black]{hyperref}
\usepackage{graphicx}
\usepackage{placeins}
\usepackage{dsfont}
\usepackage{fancyhdr}

\usepackage{multibib}
\newcites{supp}{Supplementary References}

\title{The Rlign Algorithm for Enhanced Electrocardiogram Analysis through R-Peak Alignment for Explainable Classification and Clustering}

\affiliation{Institute of Medical Informatics, University of Münster, Münster, Germany}{1}
\affiliation{Interdisciplinary Center for Clinical Research (IZKF) Münster, Münster, Germany}{2}
\affiliation{Clinic for Cardiology II: Electrophysiology, University Hospital Münster, Münster, Germany}{3}
\affiliation{ }{ }
\affiliation{These authors contributed equally to this work.}{*}

\author{Lucas Plagwitz}{1,2,*}
\author{Lucas Bickmann}{1,*}
\author{Michael Fujarski}{1}
\author{Alexander Brenner}{1}
\author{Warnes Gobalakrishnan}{1}
\author{Lars Eckardt}{3}
\author{Antonius Büscher}{1,2,3,*}
\author{Julian Varghese}{1,*}

\begin{document}
\maketitle

\begin{abstract}
Electrocardiogram (ECG) recordings have long been vital in diagnosing different cardiac conditions. Recently, research in the field of automatic ECG processing using machine learning methods has gained importance, mainly by utilizing deep learning methods on raw ECG signals. A major advantage of models like convolutional neural networks (CNNs) is their ability to effectively process biomedical imaging or signal data. However, this strength is tempered by challenges related to their lack of explainability, the need for a large amount of training data, and the complexities involved in adapting them for unsupervised clustering tasks. In addressing these tasks, we aim to reintroduce shallow learning techniques, including support vector machines and principal components analysis, into ECG signal processing by leveraging their semi-structured, cyclic form. To this end, we developed and evaluated a transformation that effectively restructures ECG signals into a fully structured format, facilitating their subsequent analysis using shallow learning algorithms. In this study, we present this adaptive transformative approach that aligns R-peaks across all signals in a dataset and resamples the segments between R-peaks, both with and without heart rate dependencies. We illustrate the substantial benefit of this transformation for traditional analysis techniques in the areas of classification, clustering, and explainability, outperforming commercial software for median beat transformation and CNN approaches. Our approach demonstrates a significant advantage for shallow machine learning methods over CNNs, especially when dealing with limited training data. Additionally, we release a fully tested and publicly accessible code framework, providing a robust alignment pipeline to support future research, available at \url{https://github.com/imi-ms/rlign}.

\end{abstract}


\section{Introduction}

Machine learning has revolutionized the field of medical signal processing, offering unprecedented insights and advancements in the diagnosis, monitoring, and treatment of various health conditions. It enables the identification of patterns and anomalies within medical signals that may not be apparent to the human eye \cite{topol_high-performance_2019}. Among the various types of medical signals, the electrocardiogram (ECG) has garnered substantial attention due to its high availability and standardization in signal acquisition and storage, and thus the availability of large datasets. ECGs, which record the electrical activity of the heart over a time period, are pivotal in diagnosing arrhythmias, myocardial ischemia, conduction disease, and other cardiac conditions. The ability to accurately analyze ECG data is, therefore, of paramount importance in cardiovascular medicine, making it a prime candidate for the application of machine learning techniques.

Major obstacles in the automatic analysis of ECG time series are variances in heart rates and temporal displacements of QRS complexes across different recordings. Although parameters such as heart rate variability may have some prognostic implications for the individual patient \cite{mejia-mejia_pulse_2020}, the temporal dispersion of unsynchronized waveforms between different ECG recordings complicates their statistical and chronological comparison. Shallow learning methods, therefore, rely on prior feature extraction from the raw ECG signal \cite{siontis_artificial_2021}. They can only learn statistical relationships within a set of predefined features that typically represent established intervals, amplitudes, or slopes around the classical ECG components during the cardiac cycle (P-wave, QRS complex, and T-wave).

Convolutional Neural Networks (CNNs) have emerged as a powerful tool for handling ECG data, thanks to their inherent ability to learn spatial hierarchies of features directly from the raw input signals \cite{lecun_deep_2015, khan_ecg_2023}. These learned features are independent of variations in heart rate and temporal offset, making CNNs highly effective for raw ECG analysis. However, they come with significant drawbacks, too: they are computationally expensive, require large numbers of positive examples, and are associated with low explainability \cite{ching_opportunities_2018, nguyen_deep_2015, garcia-martin_estimation_2019, bickmann_post_2023}. The complexity and non-transparent nature of CNNs can obscure the clinical decision-making process, making it difficult for clinicians to understand and trust the results.

In this paper, we introduce the open-source ECG alignment algorithm Rlign designed to synchronize the temporal variations across ECG recordings. First, the relationship between beats per minute and ECG structure is elucidated, followed by the development of a heart-rate-corrected alignment strategy that specifically adjusts for variations in P-onset and T-offset. We observe that R-peak alignment enhances the performance of supervised shallow learning models like support vector machines and logistic regression, allowing them to achieve or even surpass the performance of modern CNN architectures, especially in situations with limited training data. Even more, the alignment facilitates clustering of ECG time series, overcoming the challenges posed by unaligned data, where time-domain clusters are obscured by temporal misalignments of cardiac cycles. In a final analysis, we demonstrate how Rlign can be used for improved explainability of CNNs by aggregating feature attribution maps from Integrated Gradients across all instances of a data set, instead of only reviewing individual ECGs. 

All analyses within this paper are performed on the publicly available PTB-XL ECG dataset \cite{wagner_ptb-xl_2020}, ensuring full reproducibility of our results. Furthermore, we release with this work a fully tested Python framework, optimized for speed through multiprocessing.

\section{Methods}

\subsection{R-peak alignment pipeline}
\begin{figure}[ht]
      \centering
          \includegraphics[width=1\columnwidth]{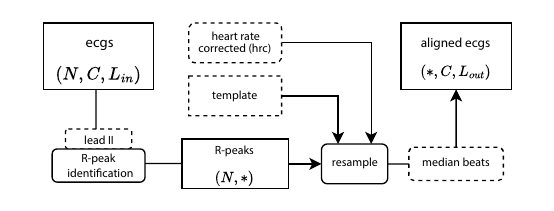}
      \caption{Schematic representation of the ECG alignment process. The diagram illustrates data in angular boxes and algorithms in rounded boxes, with dashed lines illustrating configurable settings. Key steps involve R-peak identification, correcting for heart rate variations, and resampling according to a standardized template. The final outputs are ECGs aligned at the R-peak, available as either a single median beat or the full continuous signal.}
      \label{fig:schema}
\end{figure}

The alignment comprises two main steps: initially detecting R-peaks and subsequently resampling to the form of a predetermined uniform ‘template’. The entire process is schematically visualized in Fig. \ref{fig:schema}. Our framework aligns with Scikit-Learn API conventions \cite{pedregosa_scikit-learn_2011} and features an efficient multiprocessing pipeline for rapid processing of multiple ECGs.

\subsubsection{Uniform ECG beat template}

To ensure uniformity across all ECGs, a standardized template is necessary. This template provides a consistent structure to which all signals are adjusted through transformation. It is defined by the following parameters: duration, sampling frequency, desired beats-per-minute (bpm), and an initial offset. Both the duration and sampling frequency must align with those of the input ECGs. The number of QRS complexes is influenced by the total duration of the ECG and the chosen target bpm. The initial offset determines the starting position of the first QRS complex. In summary, the template essentially specifies the locations of R-peaks and the intervals between them.

\subsubsection{R-peak detection and resampling}

R-peak detection is based on the Python package NeuroKit2 \cite{makowski_neurokit2_2021}, which offers a range of sophisticated algorithms for this purpose. As a preprocessing step, a single ECG lead (default lead II) is cleaned with a 0.5 Hz high-pass butterworth filter in fifth-order, followed by a powerline filtering in line and the detection algorithm \cite{kalidas_real-time_2017, hamilton_quantitative_1986}. We employ NeuroKit2’s default method, which identifies QRS complexes by analyzing the steepness of the absolute gradient and subsequently detects R-peaks as the local maxima within these regions.

Once R-peaks are identified and desired positions are established, the transformation process can commence. In this context, each ECG cycle is individually resampled to fit the target form through Fast-Fourier-Transform-based resampling techniques provided by the Python package scipy \cite{virtanen_scipy_2020}. Our tool provides two distinct algorithms that both align R-peaks but differ in their specific approaches. The ‘linear’-strategy takes a complete ECG cycle and transforms it into the specified unified length. A tailored ‘hrc’-strategy (short term for heart-rate-corrected) adjusts two heart-rate-related  intervals,
\begin{align*}
    \overrightarrow{P_{\text{onset}}R_{\text{peak}}}(hr)= \left[R_{peak}- \left\lfloor \frac{hr}{280}+0.14 \right\rfloor, R_{peak}\right]
\end{align*}
and 
\begin{align*}
    \overrightarrow{R_{\text{peak}}T_{\text{offset}}}(hr)= \left[R_{peak}, R_{peak}+ \left\lfloor \frac{hr}{330}+0.96 \right\rfloor \right],
\end{align*}
surrounding the R-peak from the observed heart rate to the target heart rate. Subsequently, a third resampling transformation is applied to the remaining segments. The estimation of P-onset and T-offset was conducted using the R-peak and heart rate, based on an analysis of the PTB-XL dataset.
For a detailed explanation of this phenomenon, please refer to the supplementary material \ref{sup:1}.$\;$ Finally, the aligned ECGs correspond to the specified template. Intentionally, the entire process is optimized for multiprocessing, facilitating the rapid processing of multiple ECGs.

\subsection{Evaluation and experiments}

To explore the potential of R-peak alignment as a preprocessing step, we conducted a systematic assessment of its impact on both supervised and unsupervised machine learning techniques. In the context of supervised learning, we initially evaluated performance of various R-peak alignment transformations in combination with tabular-based classifiers and with cutting-edge CNN architectures (i.e., XceptionTime \cite{rahimian_xceptiontime_2020}). Following this, we investigated the potential of explaining model decisions of both classification approaches. Furthermore, we examined the utility of R-peak aligned segments as a basis for clustering algorithms. Prior to discussing each of these aspects in detail, we begin by outlining the dataset used for our entire analysis.

\subsubsection{Simulation data}

All of our analyses are based on the PTB-XL ECG dataset \cite{wagner_ptb-xl_2020}, a large, publicly available electrocardiography dataset, in version 1.0.3. The set contains 21,799 12-lead resting-state ECGs (10s with a 500 Hz sampling rate) from 18,869 patients. All data were annotated by up to two cardiologists. Therefore, each ECG can be grouped by characteristics or abnormalities. In total, the 71 different ECG statements conform to the SCP-ECG standard and cover diagnostics. For all of our analyses, we only consider the 5 diagnostic superclasses: ‘normal ECGs’ (NORM), ‘Myocardial Infarction’ (MI), ‘Conduction Disturbance’ (CD), ‘ST-T changes’ (STTC), and ‘Hypertrophy’ (HYP). Since the dataset comes with 10 predefined folds (nine for training and one as a test set), we used this split for all our tasks. Our approach is evaluated alongside the median beats by the University of Glasgow ECG Analysis Program (Uni-G), a benchmark in commercial state-of-the-art ECG analysis packages. This data was provided by Physionet \cite{goldberger_physiobank_2000}.

\subsubsection{Classification performance and calibration}

To assess the performance of time-sensitive predictive algorithms, we evaluated two shallow learning models: logistic regression and SVM, against the advanced CNN-model, XceptionTime. These models were tested in both binary (specific abnormality vs. norm) and in the multiclass scenario, using the PTB-XL dataset superclasses. Furthermore, we conducted an in-depth analysis of the classification performance and model calibration by reviewing learning curves generated from predefined training folds, starting with one-quarter of a fold and gradually incorporating all available folds (analogous to the methods in \cite{strodthoff_deep_2021}). The analysis methodically compared the performance and calibration across both the raw time series and various forms of realigned data. Our primary performance metric was the macro-averaged area under the curve (AUC) for classification performance and expected calibration error (ECE) for mode calibration. The logistic regression and SVM models were obtained from the Scikit-Learn library, whereas the CNN models were implemented using the PyTorch-based framework tsai \cite{pedregosa_scikit-learn_2011, oguiza_tsai_2023}. The default configuration was employed for all experiments, with no optimization of hyperparameters.

\subsubsection{Model explainability}
\label{UML}
In our study, we focused not only on performance indicators but also on explainability. We evaluated global feature importance using interval-wise permutation importance \cite{plagwitz_supporting_2022}, as well as CNN-based attribution values through Integrated Gradients (IG) \cite{sundararajan_axiomatic_2017}. To compare the importance maps derived from these methods, we applied them to median beat calculation in two distinct manners.

\begin{enumerate}
    \item \textbf{Permutation Importance:} An SVM was trained using the mean-standard-deviation-scaled median beats from the pre-defined training folds of the PTB-XL dataset. We organized 19 intervals, each consisting of 25 consecutive samples. Then, we shuffled all the features within each interval and assessed the effect on the test performance compared to the original test set. This permutation process was conducted 20 times per interval. The alteration in performance was evaluated using the macro AUC, as previously utilized. A large discrepancy between original performance and the performance based on shuffled input features indicates that the respective feature group has a key influence on the decision-making process.
    \item \textbf{Integrated Gradients:} Alongside the raw signal CNN models, we conducted an analysis of the attribution maps generated by these models. Initially, a binary classification model was trained on the training set. Leveraging this model, it became feasible to compute absolute IG maps for the entirety of the test set using the Python package captum \cite{kokhlikyan_captum_2020}. IG calculates the gradient of the model's output with respect to the input features by integrating the gradients along a straight path from a baseline to the actual input. The baseline reference for the method was set to the default of zero values. Through the application of R-peak alignment, these maps were transformed from their original space to an R-peak aligned median space. This temporal alignment permitted a global aggregation of these maps by averaging across all instances.
\end{enumerate}

Due to the inherent difficulty in directly evaluating the quality of these methodologies, our analysis specifically emphasized abnormalities indicating specific regions in the ECG, notably STTC and MI (Stage $\ge$ II). We assess the correlation of all importance maps using Pearson's correlation coefficient. Higher MI stages were emphasized due to their association with distinct changes in the QRS complex, which are expected to be identifiable from maps from STTC prediction, predominantly linked to the T-wave.

\subsubsection{Unsupervised Machine Learning}

In addition to supervised learning, we explored the capabilities of Rlign in clustering algorithms. We analyzed clustering patterns across the five superclasses previously described, considering a single test fold that included 1650 ECGs. We evaluated three clustering approaches based on a 60 bpm template:
\begin{enumerate}
    \item Application of a Ward-based hierarchical clustering algorithm on a standard-scaled combined set of 12 leads (matrix of samples $\times 60,000$) \cite{ward_hierarchical_1963}.
    \item Application of the k-means algorithm on a standard-scaled combined set of 12 leads (matrix of samples $\times 60,000$).
    \item For comparative purposes, we employed a time-invariant feature extraction method called Bag-Of-Symbolic Fourier Approximation Symbols (BOSS) followed by k-means clustering on the ECG time series \cite{schafer_boss_2015}. For the multi-lead scenario, the BOSS algorithm was executed separately on each lead and subsequently combined for the k-means algorithm.
\end{enumerate}

The analysis methodically evaluated the clustering performance using both the raw time series and various forms of realigned data. Throughout these tests, we consistently used a predefined cluster count of five, in alignment with the PTB-XL superclass definitions. The performance was measured using the V-measure score, which is the harmonic mean between homogeneity and completeness, comparing the clustering results to the expert-labeled superclasses.

To demonstrate the capability of visualizing results from unsupervised analysis, we applied two versions of principal components analysis (PCA) to the data from the NORM, STTC, and MI (Stage $\ge$ II) subgroups, which are detailed in section \ref{UML}. We perform PCA on various types of data, including raw signals, Uni-G representations, Rlign linear full signals, and Rlign hrc full signals. Additionally, for the aligned data, we introduce an interval-based PCA method targeted at explainable components axes. This method involves calculating a one-dimensional PCA for both the QRS-complex and the T-wave, with specific interval start and end points predetermined by the pre-defined ECG template.

\section{Results}

\begin{figure*}[ht]
      \centering
          \includegraphics[width=2\columnwidth]{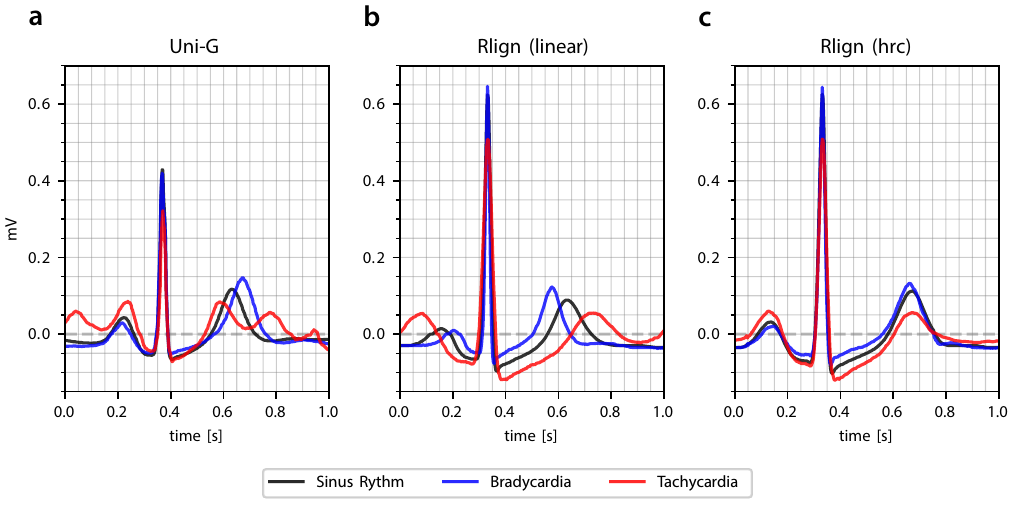}
      \caption{Summary of various median beat techniques and their impact on heart rate categories. Each chart displays three groups: sinus rhythm ($60-100$ bpm), bradycardia ($< 60$ bpm), and tachycardia ($> 100$ bpm). Panel a displays the group-specific point-wise mean beat generated using the commercial Uni-G software. This software retains the original signal sampling and does not achieve complete R-peak alignment, which is evident from the low amplitude. Panel b represents linear resampling, centered around the R-peak. Panel c illustrates our approach to resampling adjusted for heart rate, maintaining R-peak centering.}
      \label{fig:demo}
\end{figure*}

\subsection{Impact of heart rate on various resampling strategies for median beats}

Utilizing the PTB-XL ECG dataset, we computed median beats (pointwise median of all signals in one group) for ECGs categorized as normal, bradycardia, and tachycardia, to assess how different heart rate groups affect median beat representation. In Fig. \ref{fig:demo}, we compare the commercially available Uni-G algorithm (Fig. \ref{fig:demo}a) for median beat calculation with our Rlign algorithm (Fig. \ref{fig:demo}b-c). The rationale behind heart-rate-corrected (hrc) resampling stems from the observation that PQ and QT intervals, for instance, vary with heart rate, decreasing as heart rate increases. In the absence of appropriate corrections, an increasing heart rate would result in these intervals appearing disproportionately prolonged relative to the duration of the QRS complex. This phenomenon is evident in the linearly resampled median beats (Fig. \ref{fig:demo}b), where the intervals are longest for tachycardia ECGs, followed by normal ECGs, and shortest for bradycardia ECGs. Although the Uni-G algorithm incorporates some adjustments for these variations, providing satisfactory P-wave alignment, it notably results in marked displacement of T-waves. Moreover, a significant issue with Uni-G is the inclusion of more surrounding ECG signals around the single beat, potentially influencing machine learning model decisions. For example, in tachycardia ECGs, the presence of unaligned T-waves from the preceding beat and P-waves from the following beat could confound the analysis.

Conversely, our hrc resampling within the Rlign algorithm demonstrates superior P-wave and T-wave alignment (Fig. \ref{fig:demo}c). This approach ensures that the intervals are appropriately scaled relative to the QRS complex and mitigates the influence of adjacent cardiac activity on median beat calculation, regardless of the underlying ECG heart rate. 

\begin{table*}
\centering

\begin{tabular}{|c|c||c|c|c||c|c|c|c|}
\hline
\multicolumn{2}{|c||}{} & \multicolumn{3}{c||}{avg. binary} & \multicolumn{3}{c|}{multi} \\
\hline
Preprocessing & Signal & LR & SVM & CNN & LR & SVM & CNN \\ 
\hline \hline

raw signal & full & 49.2\% ($\pm$ 1.1) & 63.8\% ($\pm$ 10.1)  & \textbf{92.1\% ($\pm$ 4.5)} & 48.9\% & 62.3\%   & 87.0\% \\
\hline 
Uni-G & median  & 84.4\% ($\pm$ 9.0) & 93\% ($\pm$ 1.5) & 90.1\% ($\pm$ 6.2) & 81.3\% & 90.1\% & 88.8\% \\
\hline

\textit{Rlign} (no resampling)& median & 86.6\% ($\pm$ 7.2) & 92.6\% ($\pm$ 1.6) & 90.3\% ($\pm$ 5.7) & 82.8\% & 89.9\% & 87.6\% \\
\hline

\multirow{2}{*}{\textit{Rlign} (linear)} & median & 88.4\% ($\pm$ 4.2) & 93.3\% ($\pm$ 1.0) & 91.1\% ($\pm$ 5.0) & 85.4\% & 90.7\% & 87.8\% \\
\cline{2-8} 
 & full & 84.5\% ($\pm$ 4.8) & 92.7\% ($\pm$ 1.4) & 91.7\% ($\pm$ 4.9) & 81.1\% & 90.0\% & \textbf{89.4\%} \\
 \hline

\multirow{2}{*}{\textit{Rlign} (hrc)} & median  &  \textbf{90.1\% ($\pm$ 3.0)} & \textbf{93.4\% ($\pm$ 1.2)} & 90.8\% ($\pm$ 6.8) & \textbf{85.8\%} & \textbf{90.8\%} & 88\% \\
\cline{2-8} 
 & full & 84.9\% ($\pm$ 6.1) & 92.5\% ($\pm$ 1.5) & 90.2\% ($\pm$ 6.8) & 81.0\% & 90.0\% & 88.9\% \\ 
\hline

\end{tabular}
\caption{ECG classification performances using macro AUC for one-vs-rest scenarios across various transformation approaches and supervised learning models. Our realignment strategies are evaluated in comparison with raw data input and Uni-G median heartbeats across different classification settings. Evaluations are conducted using the diagnostic superclasses of the PTB-XL dataset. The column "avg. binary" denotes the mean performance of binary classifications, averaged across all diagnostic categories versus normal. The column "multi" signifies the overall performance in the multi-class classification setting.}
\label{tab:models_accuracy}
\end{table*}

\begin{figure*}[ht]
      \centering
          \includegraphics[width=2\columnwidth]{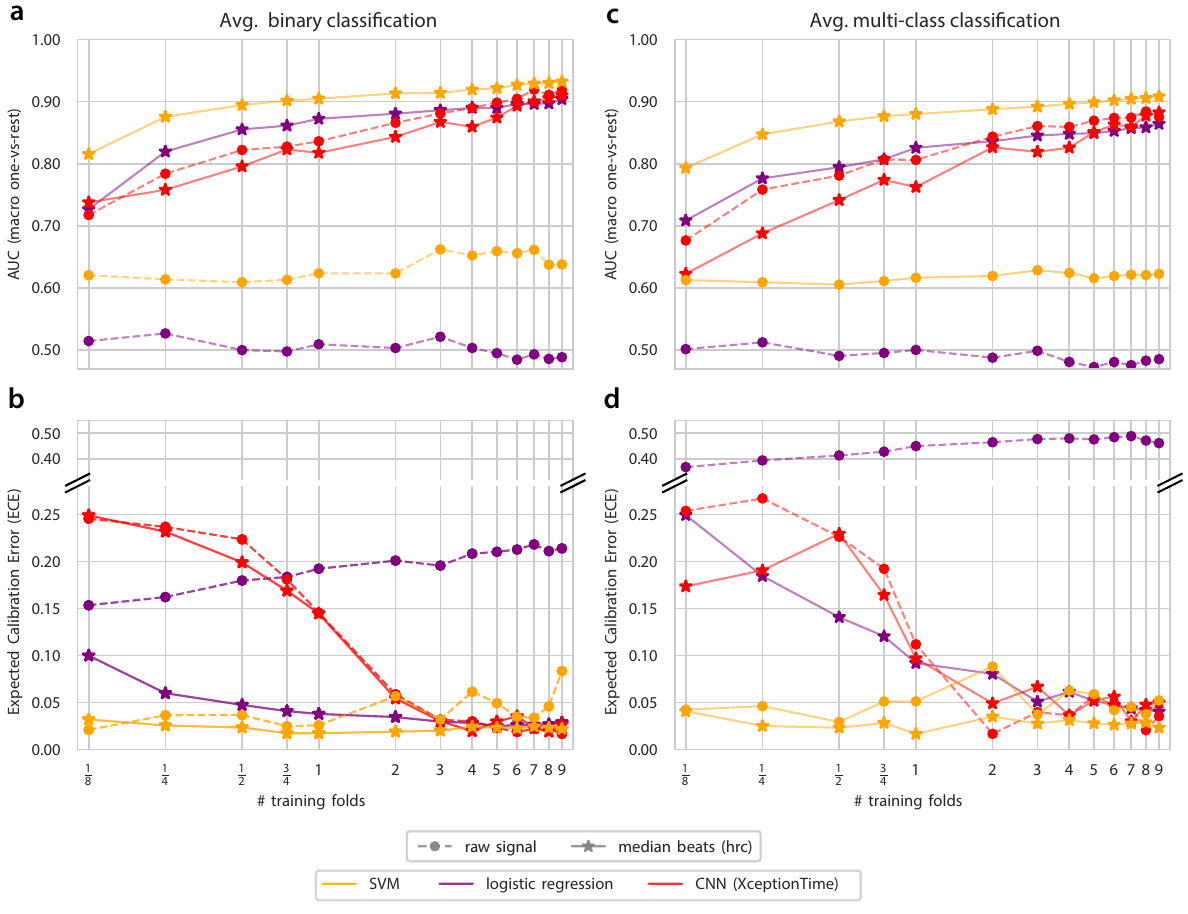}
      \caption{Classification performance and model calibration relative to the training data volume. This figure compares the effectiveness of raw signals against the implementation of median beats synchronized for heart rate variability. The comparison is made using three distinct models: linear regression, support vector machine, and the CNN architecture XceptionTime. Left column: average performance and calibration of five binary classification of abnormal vs. normal ECG; right column: overall multi-class performance and calibration. In both scenarios, a single fold consists of 1640 ECGs (Norm: 898, MI: 262, STTC: 251, CD: 174, HYP: 55).}
      \label{fig:learning_curves}
\end{figure*}

\subsection{Classification performance and calibration}
\label{subsec:classification}

To investigate the influence of R-peak alignment on classification tasks, we assessed the performance of two shallow learning models - logistic regression (LR) and support vector machine (SVM) - against a well-established multi-channel deep learning technique for time series: the CNN architecture XceptionTime \cite{rahimian_xceptiontime_2020}. Models were trained to predict the diagnostic superclasses provided in the PTB-XL dataset: ‘normal ECG’ (NORM), ‘Conduction Disturbance’ (CD), ‘Myocardial Infarction’ (MI), ‘Hypertrophy’ (HYP), and ‘ST-T changes’ (STTC). For model training, we used seven different inputs, which include: (1) raw ECG data, (2) median beats calculated by the commercial Uni-G software, (3) Rlign median-beats calculated without resampling, and (4-7) Rlign aligned ECG signals with both linear and hrc resampling techniques in both full sequence length and single median beat formats. The predictive performance of the different approaches is summarized in Table \ref{tab:models_accuracy}, with the macro area under the curve (AUC) as the primary performance metric.

Our analyses reveal that on unaligned data, only the CNN performed with reasonable performance, whereas the LR and SVM hardly surpassed the level of chance in both binary and multi-class classifications. Alignment of ECG signals with median beat calculation or resampling to aligned full-length ECGs enabled similar performances of LR and SVM as compared to the CNN, with the SVM even outperforming the CNN in most tested scenarios. Within our experimental setup, the highest predictive performance was achieved with an SVM on Rlign hrc-resampled median beats.

To further test the influence of different sample sizes on classification performance and model calibration we used the predefined cross-validation folds of the PTB-XL database. We tested 1/8, 1/4, 1/2, 3/4, 1, and multiples up to 9 folds, using the 10th fold for testing. Each single fold represented 1640 ECGs (Norm: 898, MI: 262, STTC: 251, CD: 174, HYP: 55). Our primary performance metric was the macro-averaged area under the curve (AUC) for classification performance and expected calibration error (ECE) for mode calibration. Fig. \ref{fig:learning_curves} displays the results. 

Without alignment, both LR and SVM models struggled to achieve predictive performance above the level of chance across all evaluated dataset sizes, indicating their limitations in handling unaligned data. In contrast, the CNN demonstrated predictive capabilities with unaligned data in both binary and multiclass classifications, exhibiting a near-linear relationship between predictive performance and dataset size on the logarithmic scale. However, when analyzing aligned ECG data, the SVM model displayed superior predictive performance across the entire spectrum of dataset sizes, outperforming the CNN. Similarly, LR exhibited enhanced performance with aligned data, albeit being outperformed by the CNN at larger dataset sizes. Interestingly, R-peak alignment did not improve the performance of the CNN; rather, it showed a minor trend towards decreased efficacy. This observation may reflect the CNN's capacity to internally manage unaligned data through its hierarchical feature extraction, rendering explicit alignment less beneficial and possibly introducing a form of redundancy or over-simplification for the model.

The analysis of model calibration reveals a distinct advantage for both shallow learning algorithms over the CNN when limited training data is available in binary classifications. The ECE, which quantifies the difference between predicted probabilities and actual outcomes (lower is better), for the CNN model exceeds 0.2 with half a fold of training data, whereas LR and SVM maintain an ECE below 0.05. This lower ECE indicates more reliable and accurate probability predictions from LR and SVM in the presence of limited training data. With three or more training folds, the performances of all algorithms level out. In multiclass scenarios, however, the SVM notably outperforms the others, with LR closely matching CNN's performance. In supplement \ref{sup:2}, we provide a detailed analysis of the calibration for the Norm vs. STTC classification using a single fold of training data to demonstrate the differences in model probabilities.

In summary, our findings underscore the impact of R-peak alignment on the predictive performance and calibration of machine learning models applied to ECG data analysis. While shallow models like SVM and LR significantly benefit from alignment, especially in contexts of limited data, CNNs maintain a consistent performance level, with slight variations influenced by data alignment. 

\subsection{Effects of R-peak alignment on unsupervised machine learning}

We further explored the impact of R-peak alignment on clustering, a type of unsupervised machine learning, aiming to identify inherent groupings within data without prior labeling. For this, we used the same data alignment scenarios as described previously in section \ref{subsec:classification}. We employed (1) hierarchical clustering, (2) k-means clustering, and (3) the feature extraction method from the BOSS algorithm, followed by k-means clustering for comparative analysis. We used the V-measure score, which quantifies the correlation of formed clusters with the PTB-XL diagnostic superclasses, as the primary performance metric. The results are summarized in Table 2.

\begin{figure*}[ht]
      \centering
          \includegraphics[width=2\columnwidth]{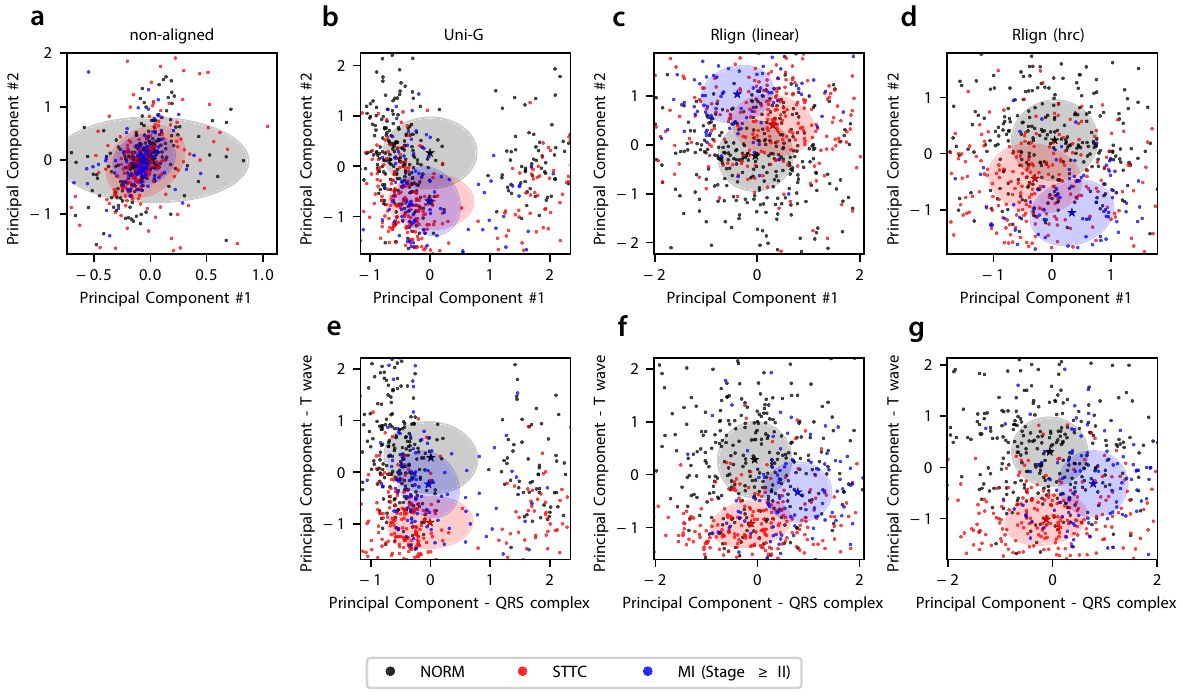}
      \caption{The realignment of ECGs enables principal component analysis (PCA) to project ECG data into a separable non-uniform two-dimensional space. The data set includes ECGs from the superclasses ‘normal ECGs’ (NORM), ‘ST-/T changes’ (STTC), and ‘Myocardial Infarction’ (MI, Stage $\ge$ II). The effect of various data transformation steps is depicted in a grid layout, with each column representing different alignment methods and each row showing different feature subsections. In panels a-d, the full time series of all 12 leads were analyzed, whereas panels e-g demonstrate the computations of two one-dimensional PCAs on the 12-lead QRS complex and T-wave separately. }
      \label{figure:pca}
\end{figure*}

Attempts to apply clustering directly on raw ECG data resulted in essentially random cluster formation, with V-measure scores of $0.7\%$ for hierarchical clustering and $0.9\%$ for k-means clustering. These scores indicate a negligible correlation between the aggregated clusters and the annotated diagnostic superclasses, underscoring the challenges of interpreting unprocessed ECG signals through unsupervised methods. However, when ECG signals were aligned using either median beats or full-length signal alignments, we observed formation of clusters that demonstrated substantial correlations with the diagnostic superclasses. Among the tested scenarios, the application of k-means clustering on hrc-resampled full ECG signals emerged as the most effective, achieving a V-measure score of $15\%$. This performance represents a significant improvement over clustering on raw data and highlights the potential of R-peak alignment in enhancing the interpretation of ECG data through unsupervised learning methods.

In comparison, BOSS identified heart-rate-independent features leading to some degree of meaningful cluster formation (V-measure score of $8.8\%$), albeit correlation to the PTB-XL superclasses was less pronounced than that achieved through direct R-peak alignment and subsequent k-means clustering. Feeding aligned median beats or full-length ECG signals into BOSS resulted in a deterioration of clustering performance compared to using raw data. This mirrors the pattern observed in the classification performance analysis, where CNNs performed best with raw data. It appears that the inherent feature extraction capabilities of methods like CNNs and BOSS are most effective when applied to unprocessed ECG data. Preprocessing, such as R-peak alignment, while beneficial for shallow learning models, may introduce biases or simplifications that adversely affect the performance of advanced feature extraction algorithms.

\begin{table*}
\centering

\begin{tabular}{|c|c||c|c|c|}
\hline
Preprocessing & Signal & Scaler $\rightarrow$ Hiearchical & Scaler $\rightarrow$ K-Means & BOSS $\rightarrow$ K-Means \\
\hline
raw signal & full & 0.7 \% & 0.9 \% & \textbf{8.8 \%}  \\ 
\hline
Uni-G & median & 7.3 \% & 9.2 \% & 2.2 \%  \\ 
\hline
\textit{Rlign} (no resampling) & median & 9.1 \% & 10.6 \% & 5.1 \%  \\ 
\hline
\multirow{2}{*}{\textit{Rlign} (linear)} & median & \textbf{13.7} \% & 11.4 \% & 5.4 \%  \\
\cline{2-5}  & full & 9.4 \% & 13.3 \% & 5.8 \%  \\ 
\hline
\multirow{2}{*}{\textit{Rlign} (linear)} & median & 10.5 \% & 13 \% & 5.2 \%  \\ 
\cline{2-5}
 & full & 10.4 \% & \textbf{15 \%} & 5.9 \%  \\ 
\hline

\end{tabular}
\caption{Comparison of clustering effectiveness across different strategies for aligned and non-aligned ECGs, as reflected by the average V-Measure score across all 12 leads. This score, which benchmarks clustering outcomes against expert-labeled superclasses, serves as an indicator of performance quality, with higher scores denoting superior clustering accuracy.}
\label{tab:models_accuracy}
\end{table*}

To illustrate the challenges posed by unsupervised learning on unaligned ECG data, we conducted a principal component analysis (PCA) for dimensionality reduction of the data while preserving as much of the data's variability as possible. Our focus was on distinguishing between the diagnostic superclasses ‘normal ECGs’ (NORM), ‘ST-T changes’ (STTC), and ‘Myocardial Infarction’ (MI, stage $\ge$ II). We excluded acute MI (stage I) as it is typically characterized by ST changes, which could overlap with the STTC class. In contrast, older MI stages typically manifest distinct alterations in the QRS complex, providing a more unique feature set for differentiation from STTC.

The PCA performed on unaligned raw data resulted in a complete overlap of the superclasses, as depicted in Fig. \ref{figure:pca}a. This overlap signifies that PCA, without prior alignment of the ECG data, fails to represent features representative for the NORM, STTC, and MI superclasses. The distribution of all three classes around a common mean suggests that the variance captured by the principal components does not correlate with the clinically relevant differences among these diagnostic classes. When applying PCA to median beats calculated by the Uni-G software, a separation between the NORM superclass and the pathological conditions (MI and STTC) begins to emerge, as shown in Fig. \ref{figure:pca}b. However, MI and STTC still completely overlap, highlighting the limitations of Uni-G in achieving a uniform alignment across the entire heart cycle. This indicates that while some degree of separation from normal ECGs is possible, the critical distinctions between MI and STTC are not adequately captured. In contrast, with linear or hrc-resampled signals processed through the Rlign algorithm, all superclasses diverge from one another in their principal components, as illustrated in Fig. \ref{figure:pca}c-d. This could explain why k-means clustering on the aligned data led to spontaneous cluster formation with some correlation to the annotated superclasses. 

Further analysis focused on differentiating the principal components associated with distinct compartments of the ECG, specifically the QRS complex and the T-wave, illustrating how R-peak alignment facilitates a more nuanced interpretation of ECG data. For Uni-G median beats (Fig. \ref{figure:pca}e), as expected, the normal ECGs and ST-T changes primarily differ in the T-wave component. However, the QRS component, crucial for identifying MI, fails to show differentiation, resulting in an overlap across all superclasses in this dimension. With Rlign processed signals (Fig. \ref{figure:pca}f-g), not only do normal ECGs and ST-T changes diverge in the T-wave component, but MI also shows a distinct pattern. Unlike ST-T changes, MI diverges less in the T-wave component but clearly in the QRS component, aligning with clinical expectations. This analysis highlights the advantage of R-peak alignment for enhancing the explainability of unsupervised learning in ECG analysis. By aligning ECG signals, Rlign allows for a clear differentiation of principal components associated with different ECG compartments and their distribution across diagnostic superclasses. 

\begin{figure*}[ht]
      \centering
          \includegraphics[width=2\columnwidth]{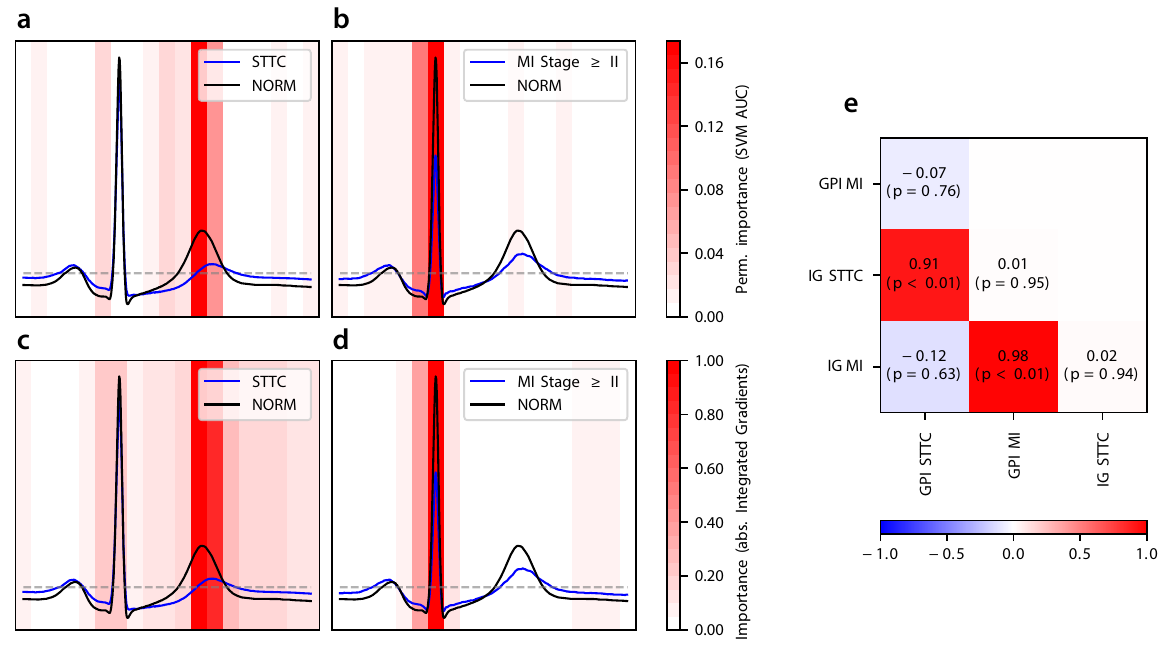}
      \caption{Importance maps and correlation for binary classification in STTC vs. NORM and MI (Stage $\ge$ II) vs. NORM. Left subfigure: The upper row illustrates the use of a Grouped Permutation Importance (GPI) method, wherein 19 intervals undergo random permutation to assess their effect on SVM prediction in terms of AUC. The lower row displays feature attribution maps obtained from the absolute values of Integrated Gradients (IG) using an XceptionTime CNN architecture. This architecture underwent training and examination on the unprocessed signal (limited to lead II), with the attribution scores aligned during a subsequent processing phase. Intervals with highlighted colors are crucial for the overall prediction. Colormaps are depicted per row. Right subfigure: The correlation matrix visualizes the correlation across GPI and IG for NORM vs. STTC and NORM vs. MI (Stage $\ge$ II).}
      \label{figure:explainability}
\end{figure*}

\subsection{Model explainability}

To further demonstrate the advantages of R-peak alignment to improve the explainability of machine learning models for ECG analysis, we conducted two exemplary binary classification tasks: ‘Normal ECG’ vs. ‘Myocardial Infarction’ (MI, stage $\ge$ II) and ‘normal ECG’ vs. ‘ST-T Changes’ (STTC). For each task, both an SVM and a CNN were trained. The relationship between input signal and output prediction was evaluated for both architectures using importance maps. Fig. \ref{figure:explainability}a-b illustrates the local permutation importance, where various intervals of the ECG signal undergo random permutation. This process assesses their impact on the SVM's predictions, with the area under the curve (AUC) serving as the performance metric. For STTC prediction, the permutation importance peaks in the T-wave area, aligning with clinical expectations that ST-T changes are predominantly reflected in this segment. Similarly, for MI (stage $\ge$ II), the highest importance is observed around the QRS complex, correlating with clinical expectations that later stages of MI exhibit distinct QRS changes on surface ECGs. Fig. \ref{figure:explainability}c-d represents a similar analysis for the CNN, employing integrated gradients (IG) to gauge signal importance. Notably, the CNN, despite being trained on unprocessed raw data, prioritizes the same ECG compartments for prediction as the SVM trained on Rlign aligned signals. The T-wave region is deemed most critical for STTC prediction, while the QRS complex is highlighted for MI, underscoring the models' alignment with clinical insights. This qualitative consistency in importance across different models can be quantitatively validated through correlation analysis (Fig. \ref{figure:explainability} right). A high correlation is observed between the IGs for STTC prediction and the global permutation importance (GPI) for STTC, but not with the GPI or IGs for MI predictions. Conversely, both importance metrics for MI show a high correlation, but not with the importances related to STTC. This analysis further reinforces the reliability and clinical relevance of the importance metrics derived from both SVM and CNN models.

A notable innovation in our approach is the application of R-peak alignment to enable aggregated importance analyses over an entire dataset for CNNs. Typically, Integrated Gradients (IG) are calculated only for individual ECGs, limiting global explainability. By initially calculating individual IG maps for the entire test set and subsequently transforming these maps from their original space to the R-peak aligned space, we facilitate global aggregation of importance metrics across all dataset instances. This process underscores the utility of R-peak alignment in enhancing model explainability by providing a novel method for interpreting CNN decisions on a dataset-wide scale.

\subsection{Acceleration through multiprocessing}

\begin{figure}[ht]
      \centering
          \includegraphics[width=.8\columnwidth]{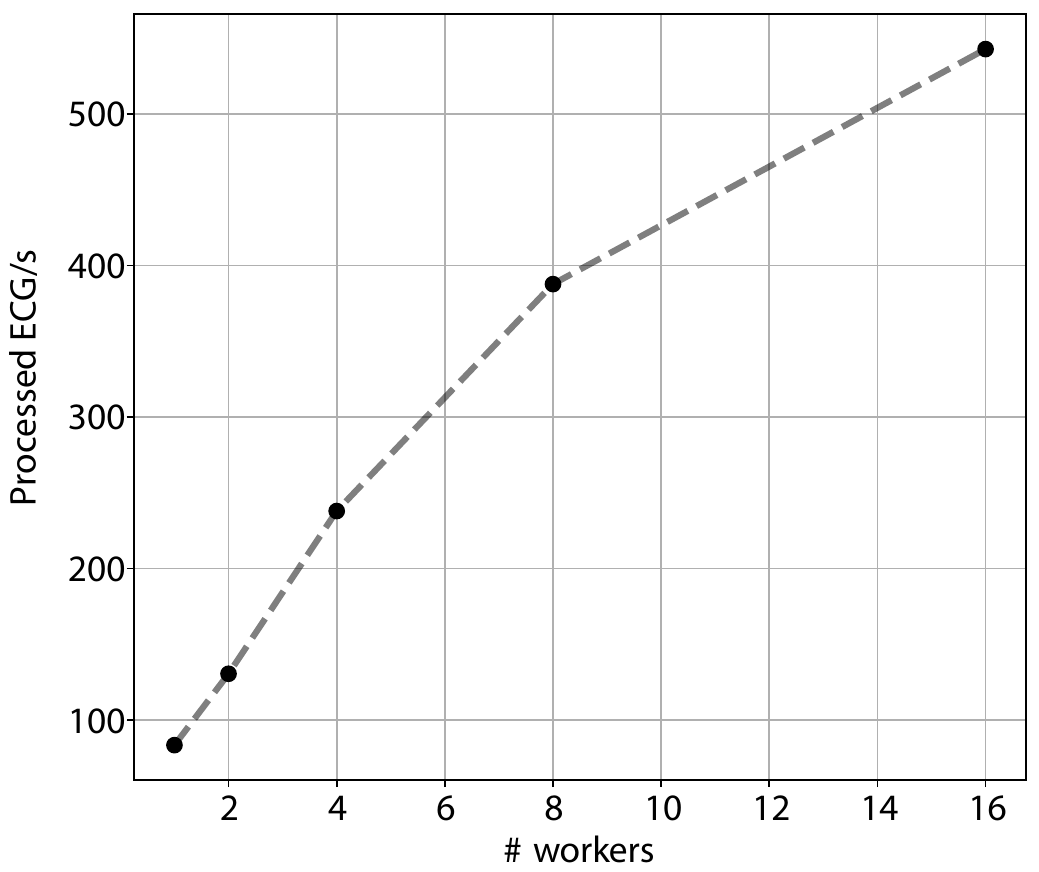}
      \caption{Acceleration through multiprocessing. This graph displays the reduction in alignment transformation time as the number of parallel workers increases from single to 2, 4, 8, 16. These tests were performed with Python 3.12 on a AMD EPYC 7543 32-core processor, utilizing resting-state ECGs (12 leads, 10 seconds, at a sampling rate of 500) as the input data.}
      \label{fig:mutliprocessing}
\end{figure}

The use of multiprocessing within the R-peak alignment pipeline significantly reduces processing times. Fig. \ref{fig:mutliprocessing} delineates the scalability of our algorithm, showcasing a reduction in processing time as the number of workers increases.

\section{Discussion}

The integration of ML models with ECG data analysis has shown promising results in enhancing the accuracy and efficiency of cardiovascular disease diagnosis. This study introduces the novel open-source ECG alignment algorithm Rlign, which synchronizes temporal variations across ECG recordings. The utilization of Rlign for R-peak alignment addresses a critical challenge in ECG analysis - the variance in heart rates and temporal displacements of QRS complexes. By aligning ECG recordings, our approach enables the application of feature-based ML models, such as SVM and LR, which traditionally lag behind deep learning models like CNNs in handling raw, unprocessed ECG data. The alignment process facilitated by Rlign not only enhances the performance of classification tasks but also contributes to their calibration and explainability. This is crucial in clinical settings, where trust in, and understanding of the rationale behind a diagnostic prediction is as important as the prediction's accuracy itself.

Our results show that tabular models, when using R-peak alignment, perform better in classification and calibration than convolutional neural networks (CNNs) when there is only a limited amount of training data available. This is a significant finding, as acquiring a sufficient amount of training cases is often challenging, especially for certain rare diseases. However, even with a sufficient amount of training data, the SVM demonstrates comparable to slightly better performance than the CNN. This suggests that, for specific diagnostic tasks within ECG analysis, simpler ML models can achieve comparable or superior classification performance to CNNs, provided that the input data is appropriately preprocessed and aligned. 

Furthermore, the impact of R-peak alignment on clustering and unsupervised learning tasks highlights the importance of preprocessing in uncovering inherent data structures obscured by temporal misalignments. The improved outcomes of unsupervised algorithms such as PCA and k-means clustering on aligned data highlight the potential of R-peak alignment to enhance ECG data interpretation and identify clinically relevant patterns without the need for prior labeling.

In terms of clinical application, the enhanced interpretability of ML models achieved through R-peak alignment is a crucial and perhaps the most relevant advancement. The ability to aggregate importance metrics across an entire dataset, as demonstrated with CNNs, offers a new avenue for understanding model decisions on a broader scale. This capability is particularly valuable in a clinical context, where it can provide insights into the model's diagnostic reasoning, potentially uncovering novel diagnostic markers or patterns.

Despite these advantages, our study also acknowledges the limitations and challenges associated with the presented preprocessing of ECG data. For instance, while R-peak alignment improves the performance of shallow models, it does not significantly enhance the performance of CNNs, even when training data is limited. This observation may suggest that CNNs' inherent feature extraction capabilities can sometimes negate the benefits of explicit alignment, indicating that the optimal preprocessing strategy may vary depending on the ML model being used. A critical aspect of ECG alignment is the potential for introducing bias into the signal that is not inherent to the data. Additionally, the alignment process or median beat calculation may filter out crucial information, such as variations between heartbeats or the presence of premature ventricular contractions (PVCs), which can be essential for certain diagnostic applications. However, our tests using the PTB-XL superclasses did not reveal this issue.

In general, the necessity of suitable time-based alignment strategies or time-invariant methods to enable biomedical signal analysis is well established. In the past, various strategies have been deployed to address the challenges posed by temporal variations within ECG recordings. Historically, alignment strategies have primarily focused on the temporal positioning of the QRS complex, given its significance in representing ventricular depolarization and its prominence in the ECG waveform. Early methods, as described by Escalona et al. (1993) \cite{escalona_fast_1993}, introduced QRS alignment through fixed signal points within a bandpass-filtered segment of the ECG, aiming to enhance the signal's clarity for subsequent high-frequency analysis. This approach demonstrated the fundamental advantage of alignment in improving the signal-to-noise ratio and facilitating more accurate detection of cardiac events. Further research by Shaw and Savard (1995) \cite{shaw_detection_1995} expanded on this concept by employing detailed alignment methods to study beat-to-beat variations. Their work highlighted the critical role of precise alignment in enabling the detailed analysis of ECG signal variations for diagnosing and monitoring cardiac abnormalities. The impact of alignment strategies on the application of deep neural networks has been shown by Xu et al. (2019) \cite{xu_towards_2019}, who demonstrated that pre-aligned data could improve network performance by up to $10\%$. However, our study did not reproduce the finding of improved CNN accuracy with pre-aligned data. This discrepancy may be due to differences in alignment strategies as well as the use of a more complex network architecture, which is better designed to mitigate overfitting and might have influenced the results.

Moreover, the use of ECG alignment for visualizing abnormalities has been a critical application, aiding clinicians in the interpretation of median beats and facilitating a more intuitive understanding of patient-specific cardiac activity. The work of Al-Zaiti et al. (2023) \cite{al-zaiti_machine_2023} refers to the use of commercial software like the Philips DXL diagnostic algorithm for alignment and feature extraction, indicating the clinical relevance and utility of these techniques. However, the proprietary nature of such tools and the lack of open-source alternatives have posed significant barriers to accessibility and widespread adoption in research and clinical settings. Our Rlign algorithm emerges in this context as a novel and accessible solution to the challenges of ECG alignment. By building upon the foundation laid by previous research, Rlign offers a robust and open-source tool for precise ECG signal alignment, facilitating the application of both traditional and machine learning-based analysis techniques. In contrast to earlier methods, Rlign is designed to be adaptable, efficient, and easily integrated into existing ECG analysis pipelines, representing a significant advancement in the field.

Another essential consideration in the application of machine learning models, particularly in the medical field, is the computational cost and power efficiency of these algorithms. Shallow learning models, such as LR, are not only computationally less demanding but also significantly more power-efficient compared to their deep learning counterparts. This difference in power consumption is critically important when deploying ML models in resource-constrained environments, such as implantable medical devices like implantable cardioverter defibrillators (ICDs) or event recorders. These devices operate under stringent power constraints, as maximizing battery life is essential to reduce the need for surgical replacements, thereby minimizing risk and discomfort for patients. The introduction of any machine learning algorithm into such devices is expected to increase computational demand, potentially shortening battery life due to the added power consumption required for processing. However, our hope with the introduction of novel preprocessing algorithms like Rlign is to mitigate the extent to which battery life is shortened. Rlign's efficiency in aligning ECG signals enables the use of less computationally demanding models for accurate ECG analysis. While the adoption of machine learning algorithms in implantable devices does present a challenge to power efficiency, the relative simplicity and lower computational requirements of these shallow models, when combined with effective preprocessing, are expected to lessen the impact on battery life compared to more complex algorithms like CNNs. For future research, we plan to evaluate Rlign with a broader range of diagnostic tests across multiple datasets to define a clearer scope. Additionally, we will benchmark the algorithm against feature-based approaches, systematically extracting predefined features in the ECG, such as specific distances and amplitudes.

Overall, Rlign emerges as a significant tool in ECG analysis. The heart-rate-corrected median beats provide a robust overview of the signal, which can be effectively interpreted by shallow learning systems such as LR and SVM. This type of signal allows the application of standard dimensionality reduction techniques and clustering methods, facilitating unlabeled pattern recognition. Moreover, CNN architectures can benefit from Rlign by aligning feature attribution maps, enhancing explainability from local to global scales. Given these capabilities, Rlign stands out as a versatile tool that not only improves the transparency but also boosts the performance of various machine learning models in ECG analysis. This opens the way for more accurate and reliable diagnostic applications.

\section*{Data \& Code availability}
All simulations are based on the publicly accessible PTB-XL data set \cite{wagner_ptb-xl_2020}. The Rlign source code is available under the terms of the MIT license at GitHub (\url{https://github.com/imi-ms/rlign}).

\section*{Conflict of Interest}
The authors declare that they have no conflict of interest.

\section*{Author contributions} 
LP, LB, AB, and JV conceptualized the work. LP and LB developed the algorithms and documentation. LP and LB contributed to the software development and maintenance. LP, LB, MF, ABr, and WG performed the experiments and analyzed the results. LP, LB, LE, AB, and JV analyzed and discussed the findings. LP and AB wrote the initial draft of the manuscript. LE and JV reviewed and edited the manuscript. AB and JV provided funding and were responsible for project administration and oversight.

\section*{Acknowledgements}

This work was supported by the Interdisciplinary Center for Clinical Research (IZKF) Münster, Germany (grant SEED/020/23 to AB).

\bibliography{main}

\clearpage
\FloatBarrier
\renewcommand{\thefigure}{S\arabic{figure}}
\section*{Supplements}
\setcounter{figure}{0}
\subsection{\label{sup:1}Heart-rate-corrected (hrc) Alignment}

\begin{figure*}[ht]
      \centering
          \includegraphics[width=1.4\columnwidth]{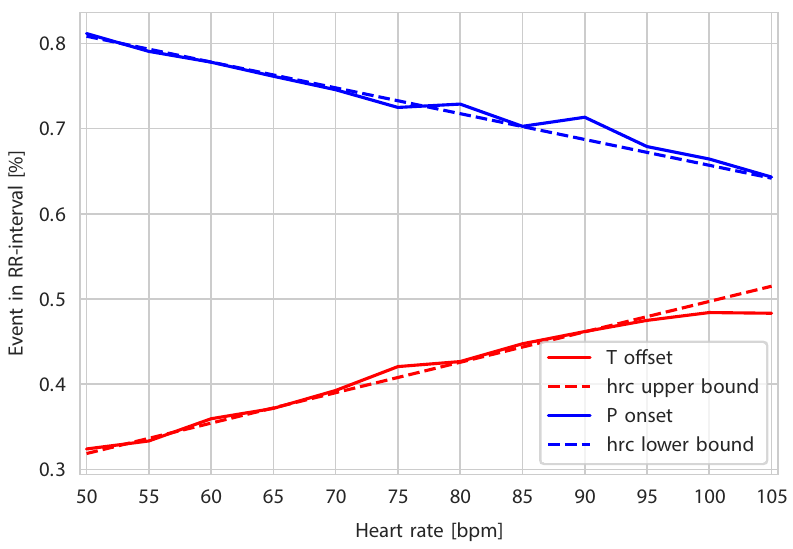}
      \caption{This figure illustrates how the P and T offsets vary within the RR interval, which changes according to heart rate. The analysis enabled the determination of interval limits for aligning ECG recordings with a specific template frequency.}
      \label{fig:supplement_hrc}
\end{figure*}

The various phases of the electrocardiogram (ECG) exhibit distinct dependencies on the underlying heart rate \citesupp{bernardo_effect_2002}. In Fig. \ref{fig:demo} of the main text, we illustrate the effectiveness of our method across bradycardia, tachycardia, and normal heart rate conditions. Our approach involves constructing a heart-rate-corrected (hrc) model that aims to accurately reconstruct the template frequency. Although existing literature explores this topic, no precise formula has yet been identified that describes the T-wave and P-wave as functions of heart rate. In further analysis, we utilized the PTB-XL dataset \cite{wagner_ptb-xl_2020}, categorizing ECGs into groups based on heart rates ranging in increments of 5 bpm steps from 50 to 105. Using Neurokit2, we computed both P-Onset and T-Offset, then normalized these measurements across the RR interval length to assess the proportion of phases within the ECG cycle. The findings of this analysis are presented in Fig. \ref{fig:supplement_hrc}. The hrc-resampling formula is a linear approximation of this. Thus, the expected event points P-Onset and T-Offset are adjusted based on the initial measured heart rate $hr_{\text{measurement}}$, the template heart rate $hr_{\text{template}}$, and the RR interval. In detail, we resample three parts of every ECG cycle with different scales. First the interval between R-peak and T-offset  

\begin{align*}
    \overrightarrow{P_{\text{onset}}R_{\text{peak}}}(hr)= \left[R_{peak}- \left\lfloor \frac{hr}{280}+0.14 \right\rfloor, R_{peak}\right]
\end{align*}
is resampled with 
\begin{align*}
\overrightarrow{P_{\text{onset}}R_{\text{peak}}}(hr_{\text{measurement}}) \mapsto \overrightarrow{P_{\text{onset}}R_{\text{peak}}}(hr_{\text{template}}).
\end{align*}

Second the interval between R-peak and T-offset 

\begin{align*}
    \overrightarrow{R_{\text{peak}}T_{\text{offset}}}(hr) = \left[R_{peak}, R_{peak} +  \left\lfloor \frac{hr}{330}+0.96\right\rfloor \right]
\end{align*}
is resampled with 
\begin{align*}
\overrightarrow{R_{\text{peak}}T_{\text{offset}}}(hr_{\text{measurement}}) \mapsto \overrightarrow{R_{\text{peak}}T_{\text{offset}}}(hr_{\text{template}}).
\end{align*}

The remaining third interval is a comparable rescaling between T-offset and P-onset of the next cycle. The hrc-resampling is therefore based on the identified R-peaks, the measured global heart rate, and the heart rate defined in the template.

\subsection{\label{sup:2}Calibration}

In the medical field, attributes such as good calibration are gaining importance alongside the pure performance of machine learning models. To evaluate a model's calibration, we follow the definition provided by Guo et al. \citesupp{guo_calibration_2017}. Specifically, we divide the possible confidence interval $[0, 1]$ into $M$ bins of size $\frac{1}{M}$. Let $B_m$ represent the set of sample indices where the prediction confidence falls within the interval $I_m = (\frac{m-1}{M}, \frac{m}{M}$]. The accuracy of $B_m$ is defined by 
\begin{align*}
 \text{acc}(B_m)= \frac{1}{B_m} \sum_{i\in B_m} \mathds{1}(\hat{y_i}=y_i),
\end{align*}
where $\hat{y_i}$ and $y_i$ are the predicted and true class labels for sample $i$. The average confidence within bin $B_m$ is defined as
\begin{align*}
\text{conf}(B_m)= \frac{1}{B_m} \sum_{i\in B_m} \hat{p_i},
\end{align*}
where $\hat{p_i}$ is the confidence for sample $i$.

\begin{figure*}[b]
      \centering
          \includegraphics[width=1.4\columnwidth]{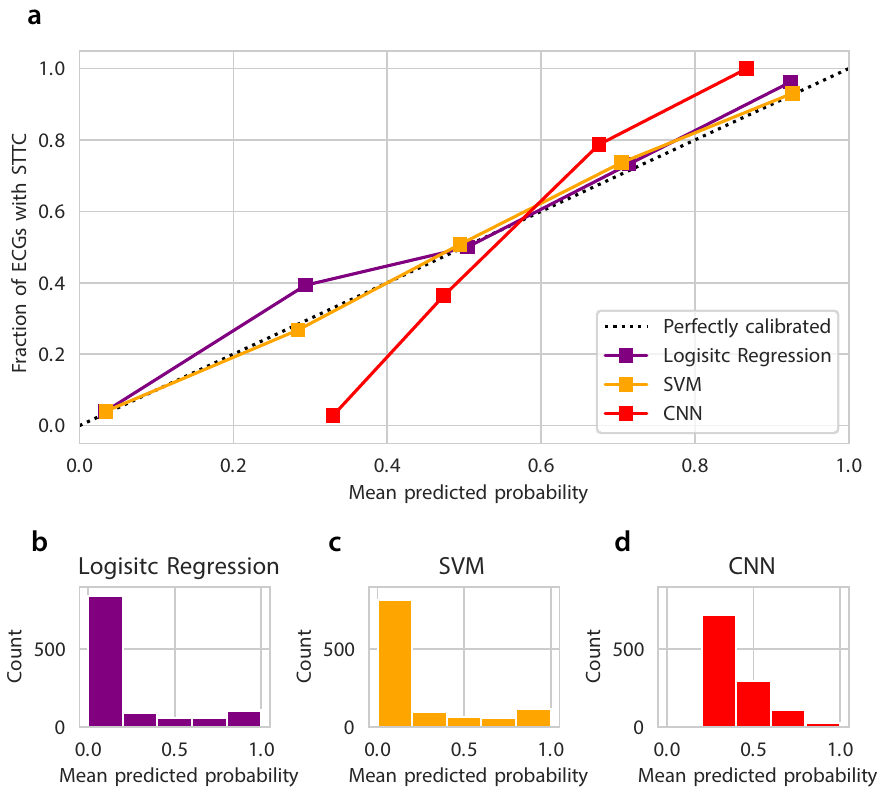}
      \caption{The calibration of the three classification algorithms (hrc-resampled median beats for logistic regression, SVM, and raw data CNN) is shown for the special case of using 1 training fold in the binary distinction between norm and STTC. All predictions were divided into 5 bins of equal width.}
      \label{fig:supp_calibration}
\end{figure*}

The expected calibration error (ECE) is given by
\begin{align*}
ECE = \sum_{m=1}^M \frac{\vert B_m \vert}{n} \left\vert \text{acc}(B_m) - \text{conf}(B_m) \right\vert, 
\end{align*}
where $n$ is the number of samples.
In the main text, the ECE loss was calculated using 10 bins, whereas Fig. \ref{fig:supp_calibration} provides a clearer depiction using 5 bins. 
Logistic regression, a calibrated method by design, is often used as a base comparison. Fig. \ref{fig:learning_curves} shows a very stable SVM through the ECE, significantly more stable than the CNN, but also calibrated as the logistic regression. With increasing training data, the ECE difference vanishes. Fig. \ref{fig:supp_calibration} shows the calibration of the models for one training fold in more detail. While the SVM is almost perfectly calibrated, the LR displays a slight shift in the second bin, where predictions in the 0.2-0.4 range are slightly overestimated. This contrasts with the CNN, which shows no prediction in the 0-0.2 range. Consequently, the first bin is entirely absent, leading to a severe underestimation in the first bin.

\bibliographystylesupp{plain}
\bibliographysupp{main}

\end{document}